 \documentclass[smallabstract,smallcaptions]{dccpaper}

\usepackage{epsfig}
\usepackage{mathtools, amsthm}
\usepackage{amsmath}
\usepackage{amssymb}
\usepackage{graphicx}
\usepackage{float}
\usepackage{booktabs}
\usepackage{array}
\usepackage{makecell}
\usepackage{multirow}
\usepackage{adjustbox}
\usepackage{subcaption}
\usepackage{algorithm}
\usepackage{algorithmic}
\usepackage{url}
\usepackage{xspace}
\usepackage[colorlinks=true, linkcolor=blue, citecolor=blue, urlcolor=blue]{hyperref}

\newlength{\figurewidth}
\newlength{\smallfigurewidth}

\newcommand{\acro}{RALHE\xspace}

\long\def\comment#1{}

\newfont{\bbb}{msbm10 scaled 700}

\newfont{\bb}{msbm10 scaled 1000}

\newcommand{\nv}{{\bf n}}

\newcommand{\vv}{{\bf v}}
\newcommand{\xv}{{\bf x}}
\newcommand{\yv}{{\bf y}}
\newcommand{\zv}{{\bf z}}

\newcommand{\Am}{{\bf A}}

\newcommand{\Cm}{{\bf C}}

\newcommand{\Dc}{{\cal D}}

\newcommand{\Lc}{{\cal L}}

\newcommand{\Qc}{{\cal Q}}

\newcommand{\Uc}{{\cal U}}

\newcommand{\muv}{\hbox{\boldmath$\mu$}}

\newcommand{\Sigmam}{\hbox{\boldmath$\Sigma$}}

\begin{document}

\title
{\large
\textbf{Region-Adaptive Learned Hierarchical Encoding for 3D Gaussian Splatting Data}
}

\author{%
Shashank N. Sridhara$^{\ast}$, Birendra Kathariya$^{\dag}$, Fangjun Pu$^{\dag}$, Peng Yin$^{\dag}$,\\ Eduardo Pavez$^{\ast}$, Antonio Ortega$^{\ast}$\\[0.5em]
{\small\begin{minipage}{\linewidth}\begin{center}
\begin{tabular}{ccc}
$^{\ast}$University of Southern California & \hspace*{0.5in} & $^{\dag}$Dolby Laboratories, Inc.\\
Los Angeles, CA, USA && Sunnyvale, CA, USA\\
%\url{{nelamang,pavezcar,aortega}@usc.edu} && \url{{birendra.kathariya, fangjun.pu, pyin}@dolby.com}
\end{tabular}
\end{center}\end{minipage}}
}

\maketitle
\thispagestyle{empty}

\begin{abstract}
We introduce Region-Adaptive Learned Hierarchical Encoding (\acro) for 3D Gaussian Splatting (3DGS) data. While 3DGS has recently become popular for novel view synthesis, the size of trained models limits its deployment in bandwidth-constrained applications such as volumetric media streaming. 
To address this, we propose a learned hierarchical latent representation that builds upon the principles of ``overfitted'' learned image compression (e.g., Cool-Chic and C3) to efficiently encode 3DGS attributes. Unlike images,  3DGS data have irregular spatial distributions of Gaussians (geometry) and consist of multiple attributes (signals) defined on the irregular geometry. 
Our codec is designed to account for these differences between images and 3DGS.
Specifically, we leverage the octree structure of the voxelized 3DGS geometry to obtain a hierarchical multi-resolution representation. Our approach overfits latents to each Gaussian attribute under a global rate constraint. These latents are decoded independently through a lightweight decoder network. To estimate the bitrate during training, we employ an autoregressive probability model that leverages octree-derived contexts from the 3D point structure. The multi-resolution latents, decoder, and autoregressive entropy coding networks are jointly optimized for each Gaussian attribute. Experiments demonstrate that the proposed \acro compression framework achieves a rendering PSNR gain of up to 2dB at low bitrates ($\leq 1$ MB) compared to the baseline 3DGS compression methods.
\end{abstract}

\section{Introduction}
3D Gaussian Splatting (3DGS) has recently emerged as a state-of-the-art approach for image-based 3D scene reconstruction \cite{kerbl2023_3dgs}. Compared to Neural Radiance Fields (NeRFs) \cite{nerf_mildenhall2021} and Plenoxels \cite{plenoxels_sara2022}, 3DGS enables much faster rendering while preserving high visual quality. In 3DGS, a scene or object is modeled explicitly as a collection of 3D Gaussians, each parameterized by a mean vector (position) and covariance matrix, along with view-dependent color and opacity as attributes \cite{kerbl2023_3dgs}. Due to its efficiency in training and rendering, 3DGS has gained rapid adoption and is expected to play a central role in future 3D content creation and immersive applications \cite{Fei2024_3dgssurvey}. Despite its efficiency, the large storage requirements for the trained 3DGS model remain a critical bottleneck in applications such as volumetric media streaming, underscoring the need for effective compression techniques \cite{Bagdasarian2024_3dgszip}.

Compression methods for 3DGS can be grouped into two categories.
First, \textit{3DGS model compaction} approaches integrate compression into model training, reducing the size of the trained 3DGS model by using techniques such as Gaussian pruning, vector quantization, and entropy-constrained optimization of Gaussian attributes \cite{lu2024_scaffoldgs, chen2024_hac3dgs, wang2024_rdogs, fan2024_lightgaussian}.
%These methods perform rate–distortion (RD) optimization independently on each 3DGS attribute, without using any spatial transforms. 
These methods do not use explicit rate-distortion (RD) optimization or any form of spatial transform, 
limiting their ability to exploit the spatial correlations of 3DGS attributes. 
In addition, these methods train an entirely new 3DGS model for each RD point. 
Second, the \textit{post-training} approaches decouple the compression pipeline from the 3DGS model training \cite{Bagdasarian2024_3dgszip}. Unlike 3DGS model compaction methods, post-training approaches do not require retraining. Instead, they encode a pre-trained 3DGS model at different bitrates. This includes encoding the attributes mapped to the 2D plane using standard video codecs \cite{lee2025_3dgsVideocomp} (for which fast implementations are available), and applying fixed transform coding tools such as geometry-based point cloud compression (GPCC) and graph Fourier transforms (GFT) \cite{xie2024_mesongs, huang2024_ptc3dgs, wang2025_adapvoxelization}. 
These methods achieve competitive RD performance with lower encoding and decoding complexities than model compaction techniques but rely on fixed, data-independent transforms, limiting their ability to fully exploit the \textit{distinct} spatial correlation of different Gaussian attributes. 
Our approach follows the post-training compression paradigm, but similar to \cite{wang2025_adapvoxelization}, includes voxelization and lightweight retraining before attribute encoding. 
The key advantage of our method is that latent representations (which play a similar role as transform domain representations in conventional methods) are  \textit{learned for each 3DGS attribute}, and thus can be adapted to their different spatial correlation characteristics, which is not possible if a fixed transform is used for all attributes. 

%Hybrid compression methods combine these ideas by adding preprocessing (e.g., voxelization) and lightweight retraining before transform coding of 3DGS attributes \cite{wang2025_adapvoxelization}. 

\begin{figure*}[t]
    \centering    \includegraphics[width=\textwidth, keepaspectratio]{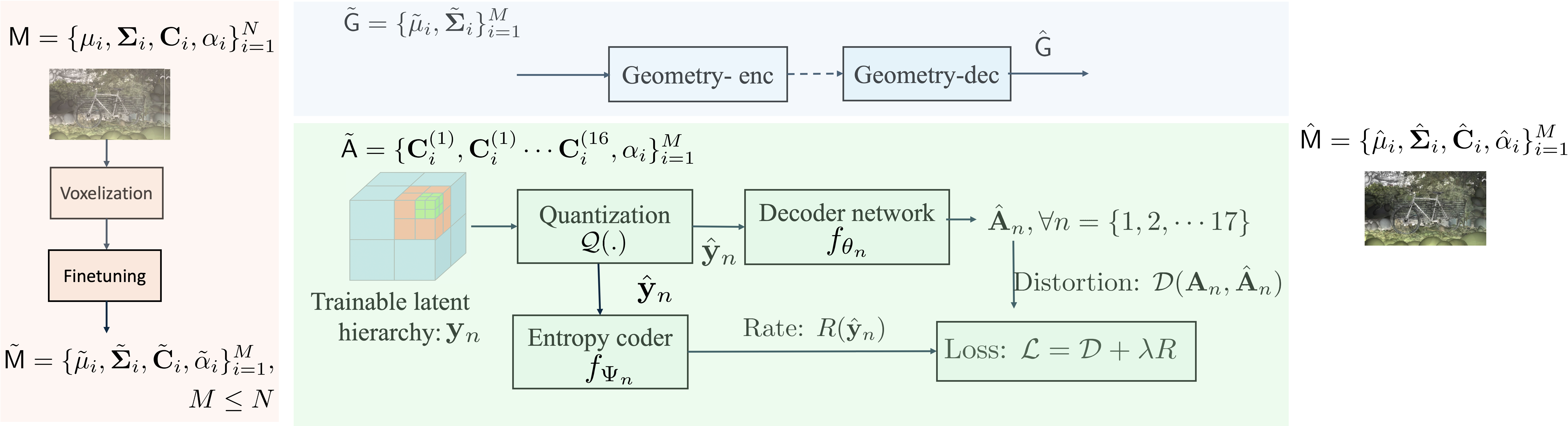}
    \caption{Overview of the proposed 3DGS compression framework. We first voxelize the Gaussian mean positions ($\muv_{i} \in \mathbb{R}^{3}$) and construct an octree to provide a hierarchical representation of the geometry. We encode the voxelized mean positions ($\tilde{\muv}_{i} \in \mathbb{R}^{3}$) using GPCC in lossless mode and covariances ($\Tilde{\Sigmam}_{i} \in \mathbb{R}^{3\times 3}$) using vector quantization. We encode the attribute data\textemdash opacity and spherical harmonic coefficient\textemdash using \acro, where we jointly train latents, decoder networks, and entropy models for each 3DGS attribute.}
    \label{fig:larch_codec_overview}
\end{figure*}

In this work, we propose a learning-based codec for efficient 3DGS compression, aiming to: 
(i) \textit{learn data-dependent latent representations} for each 3DGS attribute, rather than applying the same fixed transform (e.g., RAHT as in \cite{wang2025_adapvoxelization}) for all attributes; 
(ii) \textit{encode these representations efficiently} at different bitrates with rate-constrained optimization; 
and (iii) achieve \textit{low decoder complexity}, comparable to that of fixed transform coding methods such as GPCC.  
While autoencoder-based frameworks are widely used in learned compression \cite{balle2018_variational} and have been extended to unstructured 3D data such as point clouds \cite{sheng2022_deepPCAC} and 3DGS data \cite{chen2024_autoencoder3DGS}, their high decoder complexity limits deployment on resource-constrained devices. 
Thus, to achieve our objectives, we introduce an “overfitted” compression framework for 3DGS, inspired by learned image codecs such as Cool-Chic and C3 \cite{ladune2023_coolchic, Kim2024_c3}, which achieve low decoder complexity comparable to conventional video codecs such as HEVC. 
Our framework adopts the same principle based on jointly learning a lightweight decoder, the multi-resolution latent representation,  and an entropy coding network for the target signal.

The main challenges in extending approaches such as \cite{ladune2023_coolchic, Kim2024_c3} to 3DGS data are: 1) Gaussians are irregularly placed in 3D space, resulting in no explicit spatial ordering or regular multi-resolution representation, and 2) instead of 3 colors per pixel, 3DGS consists of 17 attributes, including direction-dependent color attributes, i.e., the spherical harmonic (SH) coefficients, as well as opacity, which exhibit distinct spatial correlations and have different effects on rendering quality.
To address the first challenge, we leverage the octree structure \cite{octree}, which offers a natural multi-resolution representation for the attributes that can be used to learn the latents. In addition,  traversing an octree following the Morton order can exploit the spatial relations embedded in the octree \cite{octree, dynamic_polygon_cloud} to provide causal contexts for the autoregressive entropy model used in rate-constrained optimization and entropy coding of latents. 
To address the second challenge, we design separate latent representations, decoders, and entropy coders for each attribute and optimize them jointly by defining a single rate-distortion function. Although our method introduces higher encoding complexity compared to fixed transform coding tools (e.g., GPCC-GS \cite{wang2025_adapvoxelization}), it achieves superior rate–distortion performance while maintaining comparable decoding complexity. 

The rest of the paper is organized as follows. 3DGS preliminaries and proposed \acro compression framework in  \autoref{sec:larch}. Experimental results and conclusions are in \autoref{sec:experiments} and \autoref{sec:conclusion}, respectively.

%\begin{figure}[th]
%    \centering
%    \includegraphics[width=0.45\textwidth, keepaspectratio]{figs/multires_latents.png}
%    \caption{Multi-resolution latent representation using octree}
%    \label{fig:multires_latent}
%\end{figure}

\section{Proposed \acro compression framework}
\label{sec:larch}
\subsection{3DGS preliminaries}
\label{sec:prelims}
3DGS data represents a scene as a set of $N$ Gaussians, each defined by a mean position $\boldsymbol{\mu}_i \in \mathbb{R}^3$, covariance matrix $\boldsymbol{\Sigma}_i \in \mathbb{R}^{3\times3}$, 16 spherical harmonic (SH) coefficients $\mathbf{C}_i \in \mathbb{R}^{16\times3}$ for view-dependent color, and opacity $\alpha_i$: $
    \mathsf{M} = \{{\muv}_i, {\Sigmam}_i, \Cm_i, \alpha_i\}_{i=1}^{N}.$
The model $\mathsf{M}$ is optimized using training images $\{\tilde{\mathbf{I}}_i\}_{i=1}^{l}$ and camera parameters $\{\tilde{\boldsymbol{\theta}}_i\}_{i=1}^{l}$. 
Quality is evaluated on test images $\{\mathbf{I}_i\}_{i=1}^{m}$ with corresponding views $\{\boldsymbol{\theta}_i\}_{i=1}^{m}$, where $R_{\boldsymbol{\theta}}$ renders the scene from view $\boldsymbol{\theta}$~\cite{kerbl2023_3dgs}. 
The training and testing losses are give by:
\begin{equation}
    \mathcal{L}_{\text{train}}(\mathsf{M}) = \frac{1}{l} \sum_{i=1}^{l} \|R_{\tilde{\boldsymbol{\theta}}_i}(\mathsf{M}) - \tilde{\mathbf{I}}_i\|_F^2, 
    \quad
    \mathcal{L}_{\text{test}}(\mathsf{M}) = \frac{1}{m} \sum_{i=1}^{m} \|R_{\boldsymbol{\theta}_i}(\mathsf{M}) - \mathbf{I}_i\|_F^2.
    \label{eq_test_distortion}
\end{equation}

\subsection{Overview of the proposed codec}
We categorize the 3DGS data into \textit{geometry}, comprising mean positions and covariances $\mathsf{G} = \{\muv_i, \Sigmam_i\}_{i=1}^{N}$, and \textit{attributes}, which consist of SH coefficients and opacities $\mathsf{A} = \{\Cm_{i}, \alpha_i \}_{i=1}^{N}$. Since the SH coefficients $\Cm_i \in \mathbb{R}^{3 \times 16}$ contain $16$ RGB triplets for an order-3 representation, there are a total of $17$ attributes.

Our proposed 3DGS compression framework consists of three stages: 
(i) preprocessing, which involves voxelization and finetuning, (ii) geometry compression, and (iii) attribute compression. Voxelization allows us to use the octree data structure to encode the positions efficiently. 
Therefore, similar to \cite{wang2025_adapvoxelization}, we first voxelize the Gaussian mean positions ($\muv_i$) to a resolution determined by the maximum octree depth $L$. 
Gaussians mapped to the same voxel are merged, yielding new positions 
$\{\tilde{\muv}_i\}_{i=1}^{M}$, where $M \leq N$. 
Following \cite{wang2025_adapvoxelization}, we perform a lightweight constrained retraining, by minimizing $\Lc_{\text{train}}$ in \eqref{eq_test_distortion} while fixing $\{\tilde{\muv}_i\}_{i=1}^{M}$, to obtain $\tilde{\mathsf{M}} = \{\tilde{\muv}_{i}, \tilde{\Sigmam}_{i}, \tilde{\Cm}_{i}, \tilde{\alpha}_{i}\}_{i=1}^{M}$, ensuring that the  rendering quality of $\mathsf{\tilde{M}}$ is close to that of $\mathsf{M}$. 
The retraining is performed only once as a pre-processing step, and the resulting model $\tilde{\mathsf{M}}$ is compressed at various bitrates. 
The voxelized positions are encoded losslessly using GPCC \cite{tmc13}. The covariances $\{\tilde{\Sigmam}_i\}_{i=1}^{M}$ are compressed using vector quantization with 
a codebook size chosen to preserve rendering quality \cite{niedermayr2024_compressed3dgs}.
%where the codebook is designed based on the relative impact of each Gaussian on the rendered images.

We denote the finetuned 3DGS attributes as  $\mathsf{\tilde{A}} = \{\tilde{\Cm}_{i}^{(1)}, \tilde{\Cm}_{i}^{(1)},\dots,\tilde{\Cm}_{i}^{(16)},\tilde{\alpha}_i\}_{i=1}^{M}$ and the matrix containing the $n$-th attribute of all points as $\Am_n$, where $n\in\{1, \ldots, 17\}$; $\Am_n$ is $M\times 3$ for $n\in\{1, \ldots, 16\}$ (there are three color components) and $\Am_{17}$ is $M\times 1$. The $i$-th row of each $\Am_n$ contains the features for the $i$-th Gaussian when scanning the points following the Morton order. We compress $\mathsf{\tilde{A}}$ using the proposed \acro. For each attribute $\Am_n$, we have  (i) latents $\yv_{n}$, (ii) an entropy model $f_{\Psi_n}$, and (iii) a decoder network $f_{\theta_n}$. Latents, entropy  models and decoder networks are jointly trained, for all attributes via backpropagation by minimizing the rate-distortion cost:
\begin{equation}
\label{eqn:overall_rdo}
    \min \sum_{n=1}^{17} \Dc(\Am_n, \hat{\Am}_n) + \lambda R(\hat{\yv}_{n}).
\end{equation}
Here $\Dc$ is the distortion between the original ($\Am_n$) and the reconstructed ($\hat{\Am}_n$) attributes, and $R$ is the estimated bitrate of the quantized latents $\hat{\yv}_{n}$. 
%It is important to note that, unlike traditional transform coding tools that first apply an analysis transform on the attributes and then entropy code the quantized transform coefficients, we directly learn the latent representation $\yv_{n}$ through optimization. 
The overview of our proposed codec is shown in \autoref{fig:larch_codec_overview}.
Overall, we train a latent representation, a decoder network, and an entropy coding network for each of the $17$ 3DGS attributes. This approach exploits the distinct spatial correlation of each 3DGS attribute effectively. In what follows, we discuss each component of \acro in detail.

\begin{figure}[t]
\centering
    \begin{subfigure}[t]{0.36\textwidth}
        \centering\includegraphics[width=\linewidth, keepaspectratio]{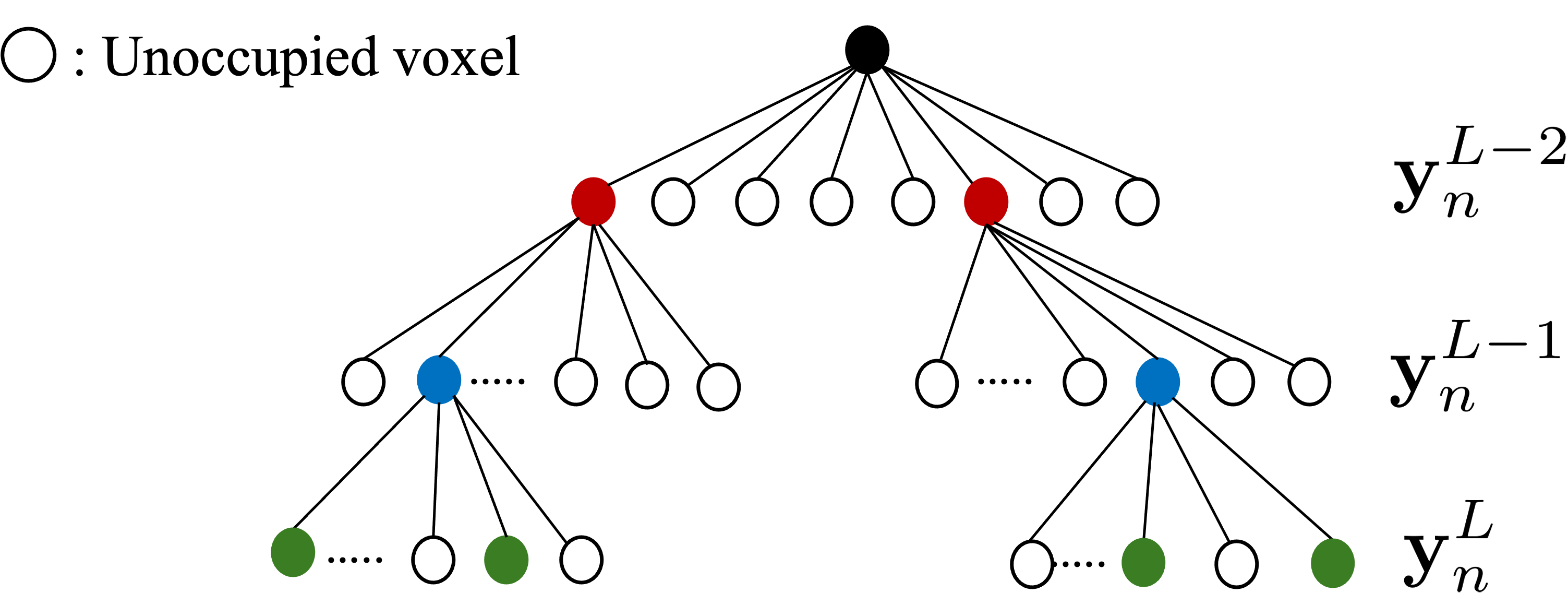}
        \caption{Multiresolution latents}\label{subfig:octree_latents}
    \end{subfigure}
    \begin{subfigure}[t]{0.63\textwidth}
    \centering
    \includegraphics[width=\linewidth, keepaspectratio]{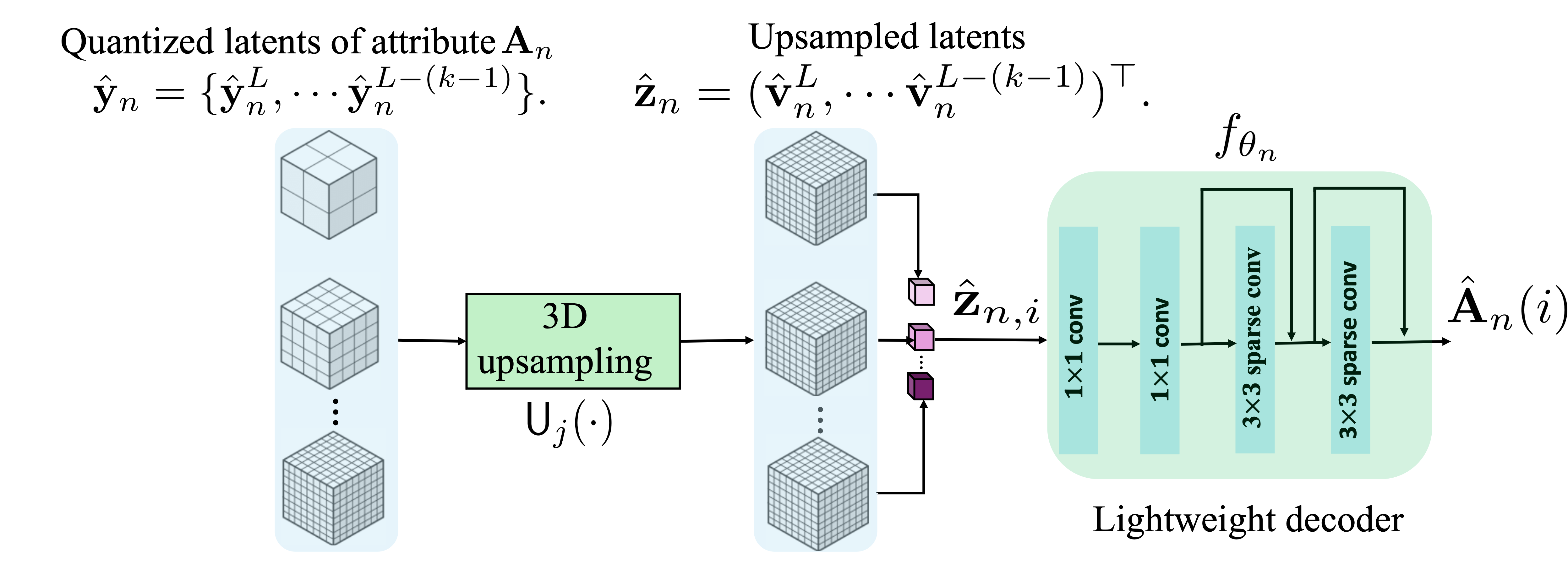}
    \caption{\acro decoder.}
    \label{subfig:synthesis}
    \end{subfigure}
    \caption{(left) hierarchical representation of 3D geometry using octree and corresponding latents, (right) reconstructing the attributes from quantized latents at different resolutions. The quantized latents are upsampled and fed into the decoder.}
    \label{fig:octree_decoder}
\end{figure}

\subsection{Multiresolution latent representation}
\label{sec:synth}
In images, multi-resolution representations are typically obtained using a pyramid structure, where coarser versions of the image are progressively produced by applying low-pass filtering followed by downsampling by a factor of two \cite{Burt1983_LaplacianPyramid, ladune2023_coolchic}. For 3DGS data, we instead exploit the octree structure, which inherently provides a hierarchical multi-resolution organization of the 3D space, as illustrated in \autoref{subfig:octree_latents}. We leverage this property to define latent representations at multiple resolutions. We consider $k$ resolutions, from the finest octree resolution $L$ to the coarse resolution $L-(k-1)$. If the bounding box of the 3D points has a volume of $W \times W \times W$, the first level of the octree divides the volume into $2^3$ cubes, each with a volume $W/2 \times W/2 \times W/2$. At each subsequent level, only the occupied cubes are divided \cite{octree, sridhara2021_cylindrical}. For the finest resolution with $L$ levels of partitioning, the resulting cubes have a volume of $W/2^L \times W/2^L \times W/2^L$, which represents the resolution of the voxelized 3D points. The coarse resolution voxels will have a volume of $W/2^{L-(k-1)} \times W/2^{L-(k-1)} \times W/2^{L-(k-1)}$. This multi-resolution depends solely on the voxelized positions $\{\tilde{\muv}_{i}\}_{i=1}^M$ and   it is the same for all attributes \cite{dynamic_polygon_cloud}. 

The matrix  $\Am_{n}$, containing the $n$th attribute for all  Gaussians, is represented by a set of $k$ latent vectors at different resolutions $\yv_{n} = \{\yv_{n}^{L}, \yv_{n}^{L-1}, \cdots \yv_{n}^{L-(k-1)}\}$,  
% \begin{equation}
%     \label{eqn:multires_latent}
%     \{\yv_{n}^{L}, \yv_{n}^{L-1}, \cdots \yv_{n}^{L-(k-1)}\},
% \end{equation}
where $\yv_{n}^{L-j}$ is the latent vector obtained from voxels at resolution $L-j$. The dimension of each latent vector depends on the number of voxels at its respective resolution. 

\subsection{\acro decoder module}
The \acro decoder module is depicted in \autoref{subfig:synthesis}. To recover  $\Am_n$, each of the quantized latent representations is upsampled to the finest resolution: $\hat{\vv}^{L-j}_{n} = \mathsf{U}_j(\hat{\yv}^{L-j}_{n})$, where $\mathsf{U}_j(\cdot)$ denotes the upsampling operation needed to go from resolution $L-j$ to resolution $L$, therefore each $\hat{\vv}_{n}^{L-j} \in \mathbb{R}^{M}$. 
The upsampled latents corresponding to the  $n$-th attribute of the $i$-th Gaussian 
$\hat{\zv}_{n,i} = (\hat{\vv}_{n}^{L}(i), \hat{\vv}_{n}^{L-1}(i), \cdots \hat{\vv}_{n}^{L-(k-1)}(i))^\top$ 
are then passed to the decoder network $f_{\theta_{n}}$ to reconstruct the target attribute $\hat{\Am}_{n}(i) = f_{\theta_{n}}(\hat{\zv}_{n,i})$. The decoder $f_{\theta_{n}}$ is a lightweight 4-layer neural network, with 3D sparse convolutions in the last two layers.
%the quantized latents at each resolution $\hat{\yv}_{n}^{L-j}$ are upsampled to the finest resolution, denoted by $\hat{\vv}_{n}^{L-j} \in \mathbb{R}^{M}$, whose $i$th entry corresponds to the $i$th Gaussian.  The upsampled latents from each resolution are passed to the decoder $f_{\theta_{n}}$.  
While image upsampling can be  carried out using bilinear or trilinear interpolation techniques  \cite{ladune2023_coolchic}, 3D upsampling  is not straightforward due to the irregular geometry. 
In this work, we adopt a simple octree-based strategy based on \textit{copying}, where in order to upsample the latents $\hat{\yv}^{L-j}_{n}$ by a factor of $2^j \times 2^j \times 2^j$, we partition the 3D points at the finest resolution into blocks of size $2^{j} \times 2^{j} \times 2^{j}$ and copy the latent values from the resolution $L-j$ to all points within the corresponding block of resolution $L$ to obtain $\hat{\vv}^{L-j}_{n} = \mathsf{U}_j(\hat{\yv}^{L-j}_{n})$. %Each of the latent representations is upsampled to the finest resolution: $\hat{\vv}^{L-j}_{n} = \mathsf{U}_j(\hat{\yv}^{L-j}_{n})$, where $\mathsf{U}_j(\cdot)$ denotes the upsampling operation needed to go from resolution $L-j$ to resolution $L$. The upsampled latents corresponding the $n$-th attribute for the $i$-th point 
%$\hat{\zv}_{n,i} = (\hat{\vv}_{n}^{L}(i), \hat{\vv}_{n}^{L-1}(i), \cdots \hat{\vv}_{n}^{L-(k-1)}(i))^\top$ 
%are then passed to the decoder network $f_{\theta_{n}}$ to reconstruct the target attribute $\hat{\Am}_{n}(i) = f_{\theta}(\hat{\zv}_{n,i})$.
% \begin{equation}
%     \hat{\Am}_{n}(i) = f_{\theta}(\hat{\vv}_{n}(i)), \text{ with } \hat{\vv}_n(i) = \{\hat{\vv}_n(i)^{L-j}, j = 0, 1 \cdots k-1\},
% \end{equation}

\subsection{Autoregressive probability model for entropy coding}
\begin{figure}[t]
    \centering
    \includegraphics[width=0.85\textwidth, keepaspectratio]{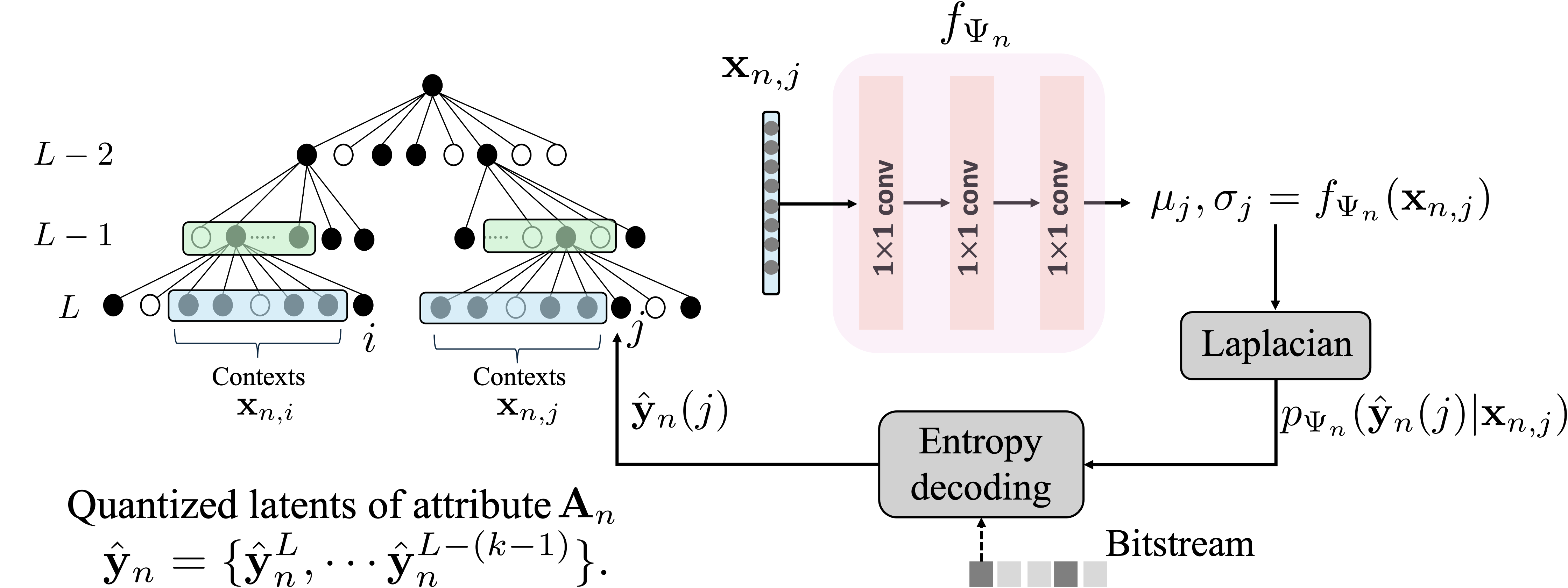}
    \caption{Autoregressive probability model for entropy coding}
    \label{fig:arm_model}
\end{figure}
The quantized hierarchical latents are encoded auto-regressively by learning a probability model $p_{\Psi_n}$ for each 3DGS attribute.  In the case of images, we typically use a rectangular context window to auto-regressively encode the pixel latents. In contrast, our method obtains causal contexts by traversing 3D space using the Morton-order (Z-order).  Morton order scanning is equivalent to performing a breadth-first traversal of the octree \cite{octree}, as illustrated in \autoref{fig:arm_model}, and it ensures that neighboring attributes are spatially close to each other in 3D space. Therefore, for a voxel $i$ in level $(L-j)$, the causal context for the attribute sorted in Morton order is given by 
$\xv_{n,i}^{L-j} = (\hat{\yv}_{n}^{L-j}(i-1), \hat{\yv}_{n}^{L-j}(i-2) \cdots \hat{\yv}_{n}^{L-j}(i-w))^{\top}$, where $w$ is the context window size.
We rely on a factorized model to obtain the joint distribution of the quantized latents $p_{\psi_{n}}(\hat{\yv}_{n})$, as widely used in learned image compression \cite{balle2018_variational, ladune2023_coolchic,Kim2024_c3}, where each factor is the distribution of the latent $\hat{\yv}_{n}(i)^{L-j}$ conditioned on its context $\xv_{n,i}^{L-j}$, thus
\begin{equation}
    p_{\psi_{n}}(\hat{\yv}_{n}) = \prod_{i, j} p_{\psi_{n}}(\hat{\yv}_{n}(i)^{L-j} |\xv_{n,i}^{L-j}).
\end{equation}
 Once the factorized probability model above is defined, we follow \cite{ladune2023_coolchic} to estimate the MLP $f_{\Psi_{n}}$. 
Finally, the rate term in \eqref{eqn:overall_rdo} is given by,
\begin{equation}
    R(\hat{\yv}_{n}) = - \log_2 p_{\Psi_{n}}(\hat{\yv}_{n}) = - \log_2 \prod_{i,j} p_{\Psi_{n}}(\hat{\yv}_{n}(i)^{L-j}|\xv_{n,i}^{L-j}).
\end{equation}
The learned entropy models $f_{\Psi_{n}}$ serve two purposes: during training, it is used to estimate and constrain the rate of the latent representation, and during encoding it is provided as parameters for entropy coding. 
\subsection{Joint optimization of RALHE for 3DGS attributes}
The optimization in \eqref{eqn:overall_rdo} can now be written as: 
\begin{equation}
    \min_{\yv_{n},\, \theta_{n},\, \Psi_{n} } \sum_{n=1}^{17} 
\Dc\!\big( \Am_{n},\; f_{\theta_{n}}(\mathsf{U}(\hat{\yv}_{n}) \big) - \lambda \sum_{n}^{17} \log_{2} p_{\Psi_{n}}(\hat{\yv}_{n}),
\end{equation}
where, $\mathsf{U}(\hat{\yv}_{n}) = [\hat{\vv}_{n}^{L}, \hat{\vv}_{n}^{L-1}, \cdots ,\hat{\vv}_{n}^{L - (k-1)}]$.
Since quantization is non differentiable, we follow standard practice and the optimization is made quantization-aware during training by adding uniform noise to the latents, $\hat{\yv}_{n} = (\yv_{n} + \nv)$, where $\nv \sim \Uc(-\frac{1}{2}, \frac{1}{2})$ as described in \cite{ladune2023_coolchic, balle2018_variational}. 
After training, the latents are uniformly quantized  resulting in $\hat{\yv}_{n} = \Qc(\yv_{n})$.
The weights and biases of the trained decoder network $\theta_{n}$ and entropy coder network $\Psi_{n}$ are quantized and encoded using an arithmetic coder. Subsequently, the quantized latents of all attributes are autoregressively encoded under a Laplace distribution, whose mean and scale parameters are estimated by the quantized entropy model $f_{\Psi_{n}}$. The same quantized entropy model is employed during both encoding and decoding of latents. 
\autoref{tab:overhead} provides a summary of all encoded quantities.
%The entropy coding network has $576$ parameters (context window of $16$), which can be sent as an overhead to the decoder.
\begin{table}[t]
\begin{center}
\caption{Details of overhead parameters that are transmitted to the decoder.}
\label{tab:overhead}
\begin{tabular}{|c|cc|cc|}
\hline
\multirow{2}{*}{} & \multicolumn{2}{c|}{decoder network $f_{\theta_{n}}$}            & \multicolumn{2}{c|}{entropy coding network $f_{\Psi_{n}}$} \\ \cline{2-5} 
                  & \multicolumn{1}{c|}{opacity:$\alpha_{i} \in \mathbb{R}$} & SH:$\Cm_{i} \in \mathbb{R}^{16 \times 3}$      & \multicolumn{1}{c|}{opacity:$\alpha_{i} \in \mathbb{R}$}        & SH:$\Cm_{i} \in \mathbb{R}^{16 \times 3}$              \\ \hline
\# parameters     & \multicolumn{1}{c|}{639}     & 639 x 16 & \multicolumn{1}{c|}{578}            & 578 x 16        \\ \hline
\end{tabular}
\end{center}
\end{table}

\section{Experiments}
\label{sec:experiments}
We evaluate the proposed \acro compression framework on 3DGS models (\textit{mic}, \textit{ficus}, and \textit{materials}) from the synthetic-NeRF dataset \cite{nerf_mildenhall2021}, each containing more than 200K Gaussians. Following \cite{wang2025_adapvoxelization}, we use uniform voxelization and lightweight fine-tuning as preprocessing steps. The voxelization depth $L_{\text{vox}}$ is selected heuristically for each model based on the point distribution and the sensitivity of rendering PSNR to quantization. Specifically, we set $L_{\text{vox}} = 14$ for \textit{mic} and \textit{materials}, and $L_{\text{vox}} = 10$ for \textit{ficus} based on the rendering quality achieved after finetuning. 
The latent representation $\yv_{n}$ is organized using the  octree into a multi-resolution hierarchy of $5$ latent grids $\{\yv_{n}^{L}, \yv_{n}^{L-1}, \ldots, \yv_{n}^{L-4)}\}$. Finally, the context window size of the autoregressive entropy coding network is fixed to $w = 16$. The RALHE framework was trained on an Nvidia RTX-2080 GPU for 10K iterations. %A detailed evaluation of RALHE on large datasets, such as MipNeRF-360, is left for future work. 

\begin{figure*}[t]
    \centering
    \begin{subfigure}[t]{0.19\textwidth}
        \centering
        \includegraphics[width=\linewidth]{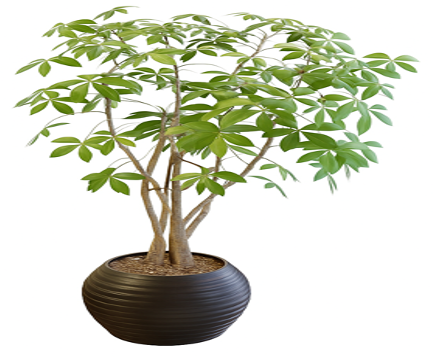}
        \captionsetup{labelformat=empty}
        \subcaption{ficus}
    \end{subfigure}
    \begin{subfigure}[t]{0.39\textwidth}
        \centering
        \begin{subfigure}[t]{0.49\textwidth}
            \centering
            \includegraphics[width=\linewidth]{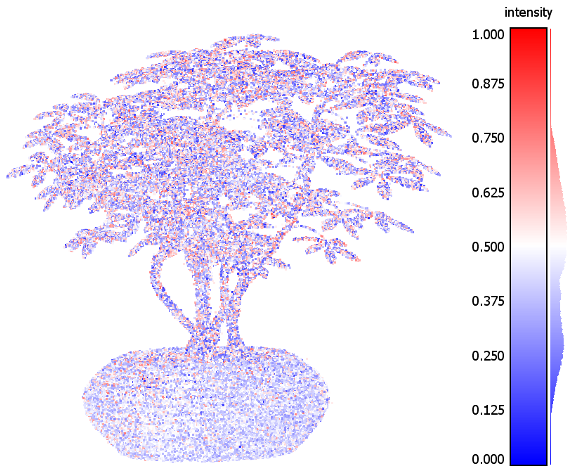}
            \captionsetup{labelformat=empty}
            \caption{Low resolution}
        \end{subfigure}%
        \hfill
        \begin{subfigure}[t]{0.49\textwidth}
            \centering
            \includegraphics[width=\linewidth]{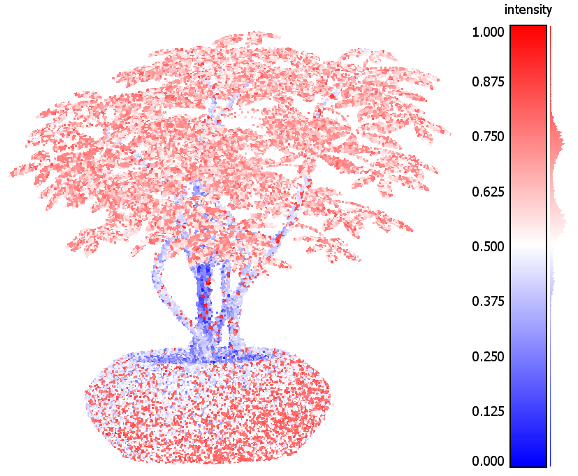}
            \captionsetup{labelformat=empty}
            \caption{High resolution}
        \end{subfigure}
        \captionsetup{labelformat=empty}
        \subcaption{(a) Low bitrate (color): $0.034$ bpp}%
    \end{subfigure}
    \hfill
    % Right block (two images + one subcaption for the block)
    \begin{subfigure}[t]{0.39\textwidth}
        \centering
        % First row: 2 subfigures inside
        \begin{subfigure}[t]{0.49\textwidth}
            \centering
            \includegraphics[width=\linewidth]{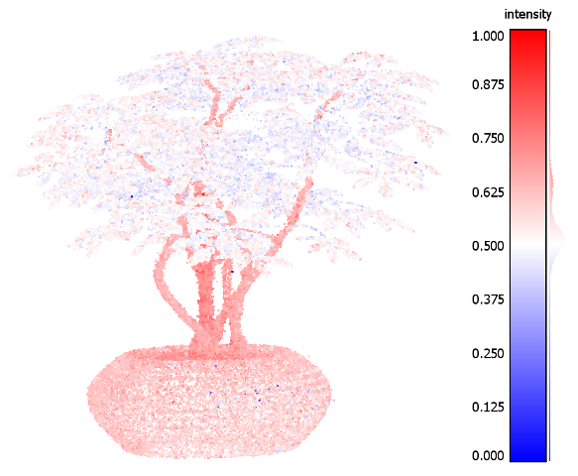}
            \captionsetup{labelformat=empty}
            \caption{Low resolution}
        \end{subfigure}%
        \hfill
        \begin{subfigure}[t]{0.49\textwidth}
            \centering  
            \includegraphics[width=\linewidth]{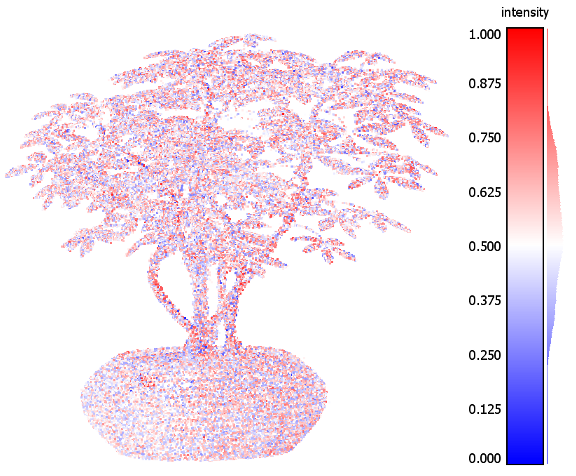}
            \captionsetup{labelformat=empty}
            \caption{High resolution}
        \end{subfigure}
        \captionsetup{labelformat=empty}
        \subcaption{(b) High bitrate (color): $2.68$bpp}
    \end{subfigure}
    \caption{Visualization of learned latents for color attributes i.e., $0^{\text{th}}$ order SH: $\Cm_{i}^{(0)}$.}
    \label{fig:visualization}
\end{figure*}

We first provide a visualization of the learned latents for the color attribute at different bitrates for the \textit{ficus} model in \autoref{fig:visualization}. For visualization, low-resolution latents are upsampled to match the full resolution. At high bitrates, the hierarchical latents exhibit a clear separation of roles: low-resolution latents capture coarse semantic structure (e.g., distinguishing leaves from the pot and branches), while high-resolution latents provide fine details. In contrast, at low bitrates, the representational capacity is limited, and even high-resolution latents primarily encode low-frequency information, resulting in the loss of fine details. 
%This demonstrates how bitrate allocation directly affects the ability of the latent hierarchy to balance global structure and fine details. 

\subsection{Comparison with state-of-the-art 3DGS compression methods}
\begin{figure}[t]
    \centering
        \begin{subfigure}[t]{0.33\textwidth}
            \centering
            \includegraphics[width=\linewidth]{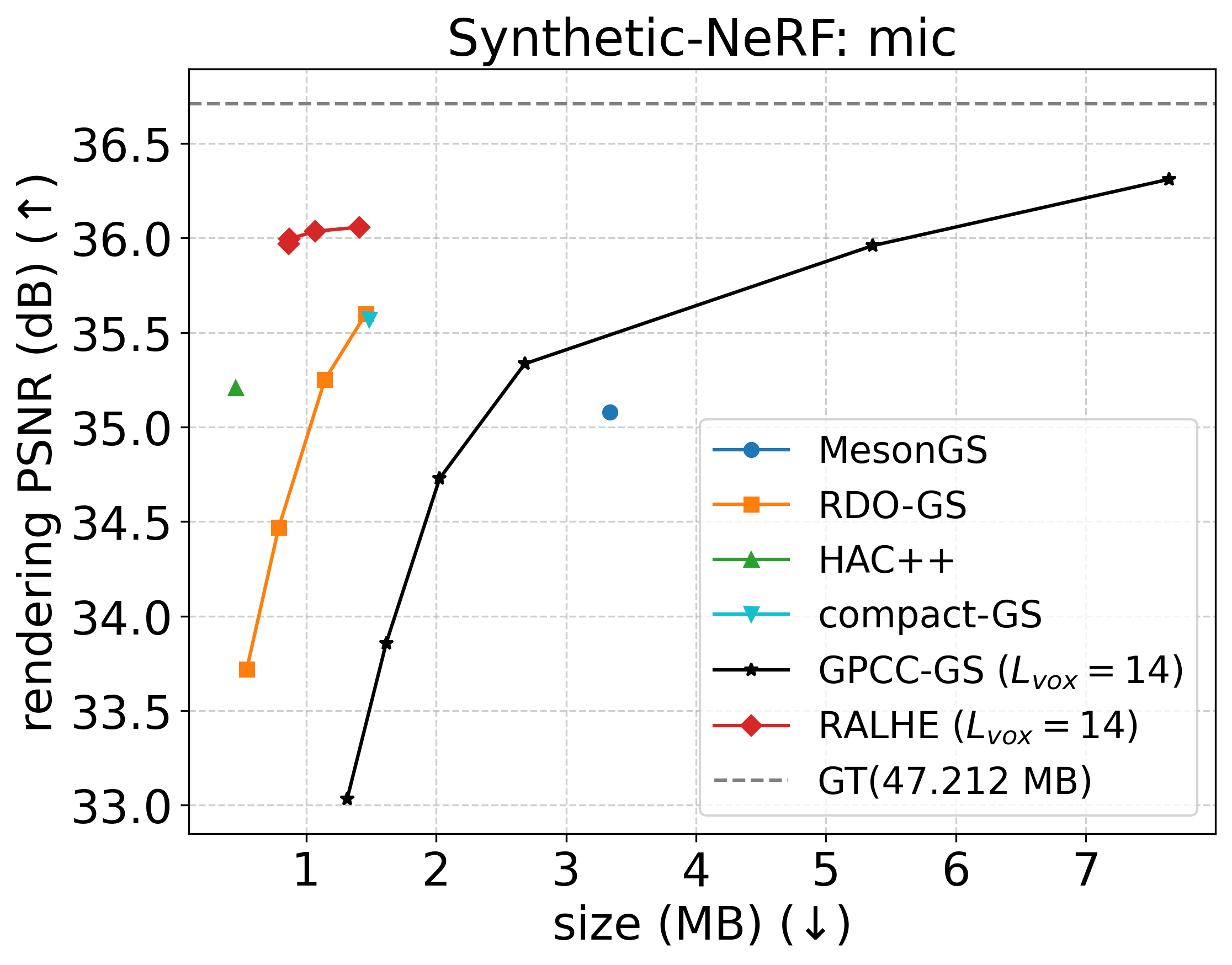}
        \end{subfigure}%
        \hfill
        \begin{subfigure}[t]{0.33\textwidth}
            \centering
            \includegraphics[width=\linewidth]{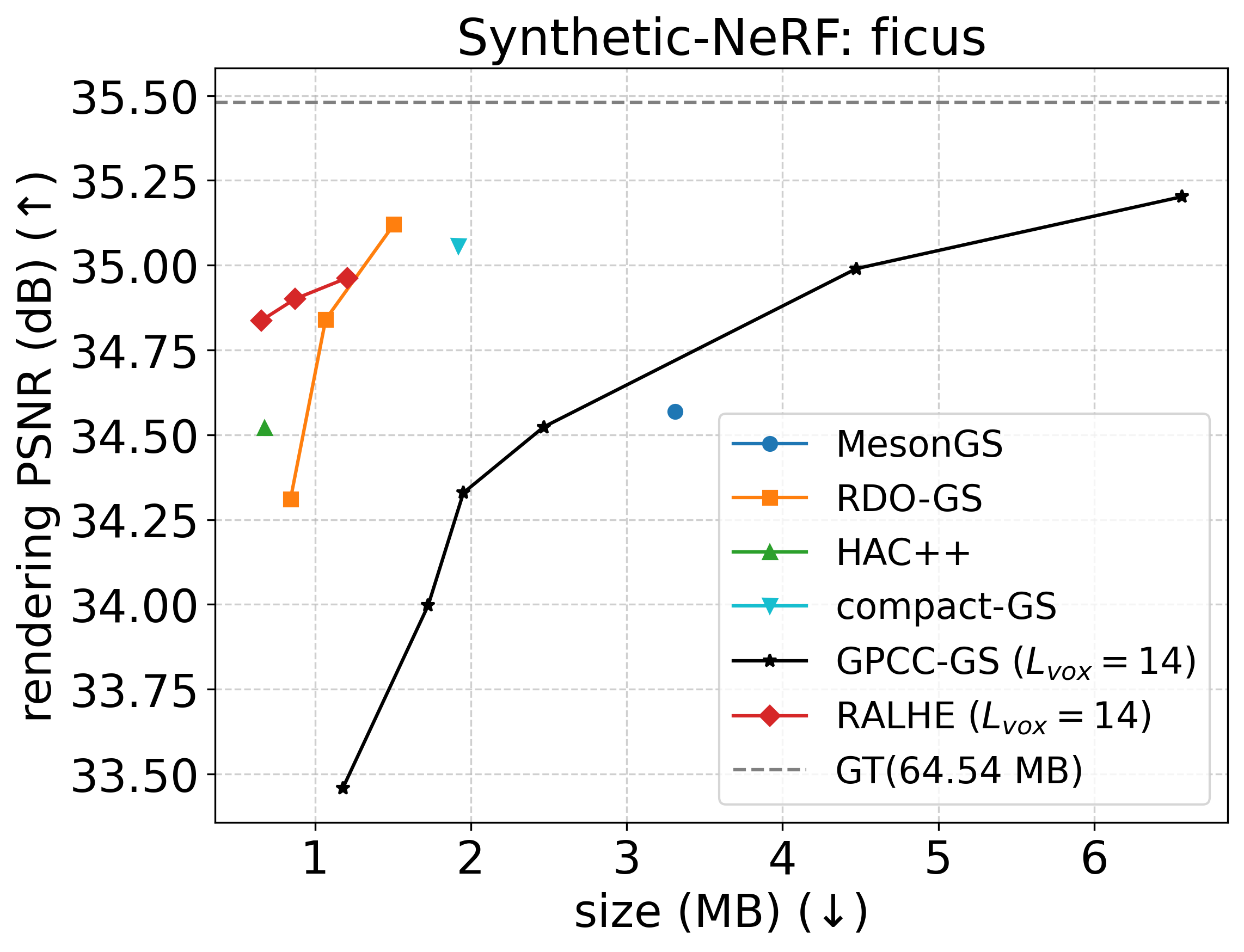}
        \end{subfigure}%
        \hfill
        \begin{subfigure}[t]{0.33\textwidth}
            \centering
            \includegraphics[width=\linewidth]{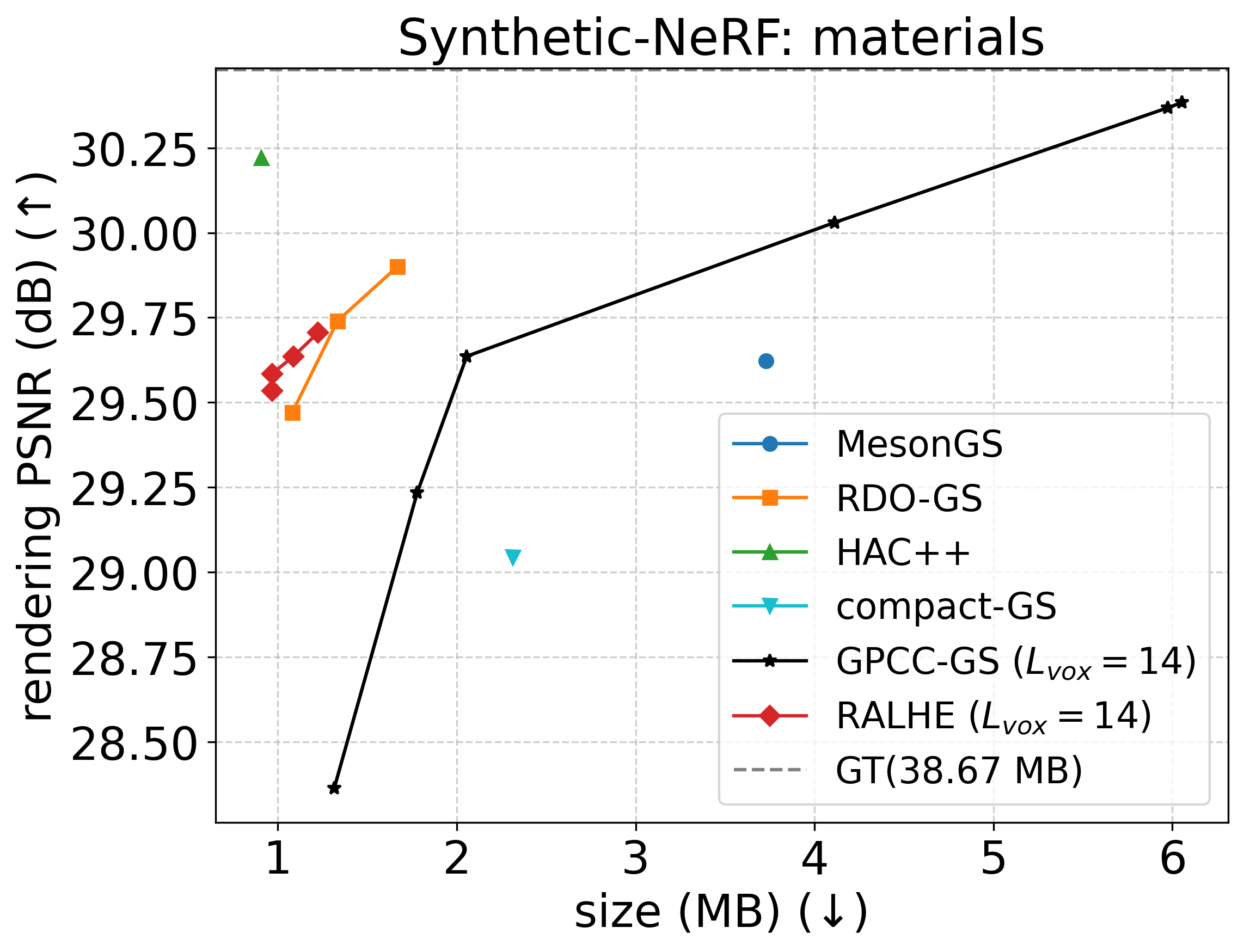}
        \end{subfigure}
        \caption{RD cuve comparison with state-of-the-art 3DGS compression methods.}%
        \label{fig:rd_comparison}
\end{figure}

We compare the proposed \acro codec with state-of-the-art 3DGS compression approaches. 
For model compaction baselines, we consider RDO-GS \cite{wang2024_rdogs}, Compact-GS \cite{navaneet2023_compgs}, and HAC++ \cite{chen2024_hac3dgs}. For post-training baselines, we include MesonGS \cite{xie2024_mesongs} and GPCC-GS with adaptive voxelization\cite{wang2025_adapvoxelization}. As shown in \autoref{fig:rd_comparison}, our proposed \acro codec outperforms the baselines for \textit{mic} and \textit{ficus} models by a large margin. For \textit{materials}, however, our method achieves R–D performance comparable to existing model compaction methods. This is because the \textit{materials} model contains multiple spatially separated objects. In such cases, the octree structure groups Gaussians from different objects into the same blocks, which reduces reconstruction accuracy and leads to less reliable bitrate estimation.

Finally, we compare the rendering PSNR of the proposed RALHE at a fixed bitrate of 1 MB with RDO-GS \cite{wang2024_rdogs} and GPCC-GS \cite{wang2025_adapvoxelization}\footnote{Due to insufficient overlap between the R–D curves, accurate B-D rate and B-D PSNR values cannot be computed.}. As shown in \autoref{tab:decoding_time}, RALHE achieves a PSNR gain of 0.3–0.7 dB over RDO-GS and up to 2.0 dB over GPCC-GS at the same bitrate. We also report the decoding time per R–D point in \autoref{tab:decoding_time}. The GPCC-GS decoder performs entropy decoding and inverse transforms of the coefficients, while RDO-GS reconstructs attributes from the codebook index. For fairness, all baseline decoders run on the CPU. RALHE’s decoding time lies between the two, with GPCC-GS being the fastest due to its optimized C++ implementation.

\section{Conclusion}
\label{sec:conclusion}
In this work, we proposed a  3DGS compression framework, where the  geometry (positions and Gaussians) is encoded first,  followed by 3DGS attribute encoding,  conditioned on the decoded geometry. Our main contribution is \acro to encode  3DGS attributes, which ``overfits'' a multi-resolution latent representation, a lightweight decoder, and an entropy coder for a given target signal. Since 3DGS consists of multiple attributes (signals), we optimize a separate latent representation, decoder, and entropy coder jointly for each attribute. We leverage the octree structure to obtain a multi-resolution representation for latents and to derive contexts for the autoregressive probability model used for entropy coding. The proposed \acro decoder reconstructs the original attribute using quantized latents that are upsampled to the finest resolution using the octree. The proposed framework, combined with existing geometry coding solutions, achieves state-of-the-art coding performance. 

%\begin{figure}[t]
%\begin{center}
%\begin{tabular}{cc}
%\multicolumn{2}{c}{\epsfig{width=4in,file=Figures/image1}} \\
%\multicolumn{2}{c}{\small{(a)}} \\[1em]
%\epsfig{width=2in,file=Figures/image3.eps} &
%\epsfig{width=2in,file=Figures/image4.eps} \\
%{\small (b)} & {\small (c)}
%\end{tabular}
%\end{center}
%\caption{\label{fig:example}%
%An example figure.}
%\end{figure}

\begin{table}[tp]
\begin{center}
\caption{Comparison of rendering PSNR at a fixed bitrate of 1MB and decoding time of RALHE with 3DGS model compaction and post-training compression method}
\label{tab:decoding_time}
{\footnotesize 
\begin{tabular}{|c|ccc|ccc|}
\hline
\multirow{2}{*}{model} & \multicolumn{3}{c|}{rendering psnr (dB) at 1MB}                            & \multicolumn{3}{c|}{decoding time (seconds)}                             \\ \cline{2-7} 
                          & \multicolumn{1}{c|}{RDO-GS\cite{wang2024_rdogs}}  & \multicolumn{1}{c|}{GPCC-GS\cite{wang2025_adapvoxelization}} & RALHE   & \multicolumn{1}{c|}{RDO-GS\cite{wang2024_rdogs}} & \multicolumn{1}{c|}{GPCC-GS\cite{wang2025_adapvoxelization}} & RALHE \\ \hline
mic                       & \multicolumn{1}{c|}{35.25} & \multicolumn{1}{c|}{32.06} & \textbf{36.03} & \multicolumn{1}{c|}{5.92}       & \multicolumn{1}{c|}{\textbf{2.38}}   & 4.78 \\ 
ficus                     & \multicolumn{1}{c|}{34.64} & \multicolumn{1}{c|}{33.58} & \textbf{34.93} & \multicolumn{1}{c|}{6.38}       & \multicolumn{1}{c|}{\textbf{3.24}}   & 5.05 \\
materials                 & \multicolumn{1}{c|}{29.47} & \multicolumn{1}{c|}{27.16} & \textbf{29.63} & \multicolumn{1}{c|}{4.85}       & \multicolumn{1}{c|}{\textbf{2.73}}   & 4.43 \\ \hline
\end{tabular}
}
\end{center}
\end{table}

\section{References}
\bibliographystyle{IEEEtran}
\bibliography{refs}

\end{document}